\documentclass[twocolumn]{aastex61}
\usepackage{epsfig}
\usepackage{epstopdf}

\received{April 20, 2017}
\revised{May 10, 2017}
\accepted{May 15, 2017}
\submitjournal{ApJ Letters}

\shorttitle{Statistical Comparative Planetology}
\shortauthors{Bean, Abbot, \& Kempton}

\begin{document}

\title{A Statistical Comparative Planetology Approach to the Hunt for Habitable Exoplanets and Life Beyond the Solar System}

\correspondingauthor{Jacob L. Bean}
\email{jbean@oddjob.uchicago.edu}

\author{Jacob L. Bean}
\affil{Department of Astronomy \& Astrophysics, University of Chicago, 5640 S. Ellis Avenue, Chicago, IL 60637, USA}
\author{Dorian S. Abbot}
\affil{Department of the Geophysical Sciences, University of Chicago, 5734 S. Ellis Avenue, Chicago, IL 60637, USA}
\author{Eliza M.-R. Kempton}
\affil{Department of Physics, Grinnell College, 1116 8th Avenue, Grinnell, IA 50112, USA}

\begin{abstract}
The search for habitable exoplanets and life beyond the Solar System is one of the most compelling scientific opportunities of our time. Nevertheless, the high cost of building facilities that can address this topic and the keen public interest in the results of such research requires the rigorous development of experiments that can deliver a definitive advance in our understanding. Most work to date in this area has focused on a ``systems science'' approach of obtaining and interpreting comprehensive data for individual planets to make statements about their habitability and the possibility that they harbor life. This strategy is challenging because of the diversity of exoplanets, both observed and expected, and the limited information that can be obtained with astronomical instruments. Here we propose a complementary approach that is based on performing surveys of key planetary characteristics and using statistical marginalization to answer broader questions than can be addressed with a small sample of objects. The fundamental principle of this comparative planetology approach is maximizing what can be learned from each type of measurement by applying it widely rather than requiring that multiple kinds of observations be brought to bear on a single object. As a proof of concept, we outline a survey of terrestrial exoplanet atmospheric water and carbon dioxide abundances that would test the habitable zone hypothesis and lead to a deeper understanding of the frequency of habitable planets. We also discuss ideas for additional surveys that could be developed to test other foundational hypotheses is this area.
\end{abstract}


\section{Introduction} \label{sec:intro}
Through the study of exoplanets, humanity stands on the threshold of making significant progress towards answering the age-old question of whether there is life elsewhere in the Universe. Exoplanet surveys, and in particular NASA's \textit{Kepler} mission, have revealed that small planets are common in circumstellar habitable zones in our Galaxy \citep{petigura13,burke15,dressing15}. The search for exoplanets has recently culminated in the discovery of the first Earth-size planets in the habitable zones of nearby stars \citep{angladaescude16,gillon17,dittmann17}. By virtue of their orbiting nearby stars, and with the pending advent of powerful new instruments and facilities, these newly discovered planets are the first bona fide targets for future efforts using the techniques of astronomical remote sensing to determine if they are truly habitable and possibly even inhabited \citep{kreidberg16,meadows16,barstow16,turbet16,lovis17}.

The exciting possibility of finding other Earth-like worlds and life beyond our Solar System has motivated the development of new instruments for existing telescopes and is a key justification for the construction of the next generation of extremely large ground-based telescopes (``ELTs''). The characterization of potentially habitable planets is also expected to be a major part of the science program for the \textit{James Webb Space Telesscope} (\textit{JWST}), which is planned for launch in 2018 \citep{deming09,beichman14,cowan15}. Furthermore, two of the four flagship mission concepts currently being developed by NASA in preparation for the next Astronomy and Astrophysics Decadal Survey have characterization of terrestrial exoplanets as a main driver\footnote{The \textit{Large UV/Optical/IR Surveyor} (\textit{LUVOIR}: \url{https://asd.gsfc.nasa.gov/luvoir/}) and the \textit{Habitable Exoplanet Imaging Mission} (\textit{HabEx}: \url{http://www.jpl.nasa.gov/habex/}).}, while a third is being designed with this science case as an option\footnote{The \textit{Origins Space Telescope} (\textit{OST}: \url{https://asd.gsfc.nasa.gov/firs/}).}.

With the fast-approaching opportunity to make a search for habitable environments and life on exoplanets comes the very real challenge of actually designing an experiment that will deliver clear results. The scientific challenges of designing experiments on this topic are formidable because terrestrial exoplanets are expected to be diverse in the structures and compositions of their interiors and atmospheres. This expectation is based on our knowledge of the current and past states of the terrestrial planets of the Solar System, the diversity of bulk compositions inferred for the known low-mass exoplanets, the different properties of the stars in the solar neighborhood, and the random nature of the planet formation and evolution processes.

The challenges of designing a robust experiment are compounded by the fact that even the best telescopes and instruments currently under construction or in design will only reveal a small piece of a planet's puzzle on their own. This is due to the practical limitations of certain approaches \citep[e.g., the challenge of determining planet masses with direct imaging,][]{brown15} and the finite grasp of instruments as set by technological constraints or fundamental physics (e.g., not having sufficient spectral coverage to detect all the chemical species of interest in an exoplanet's atmosphere using a single instrument). Therefore, the questions of how many planets need to be characterized, which planetary properties need to be determined, and what level of precision is needed in the measurements are not trivial to answer, yet they have profound implications for the cost, risk, and timescale of the program \citep{stark15,stark16}.

To answer this experimental design challenge, we suggest that a statistical comparative planetology approach should be a key element in efforts to address the topics of habitable worlds and life beyond the Solar System. Statistical comparative planetology requires a broad survey that will necessarily be less detailed than what could be obtained with an approach focused on a small number of planets. However, the fundamental principle of this approach is maximizing what can be learned from each type of measurement by applying it widely as an alternative to bringing multiple detailed measurements to bear on a single object. The advantages of this approach are that a broad survey can give context to aid the understanding of individual planets, and it enables conclusions to be reached statistically despite ambiguous results for individual objects. The statistical comparative planetology approach also uses the diversity of exoplanets as an advantage to be exploited rather than a challenge to be overcome. Additionally, this approach can be built around simple physical models instead of the more complex models needed to make accurate statements about an individual planet. 

The statistical comparative planetology approach we advocate is an extension of the hugely successful efforts to determine planetary frequency, and it is informed by lessons learned in the atmospheric characterization of close-in transiting exoplanets. However, our proposal is somewhat at odds with the prevailing ``planetary systems science'' approach to the problem. We therefore begin in \S \ref{sec:systems} with a review of the planetary systems science approach to addressing the topics of habitable worlds and life beyond the Solar System. In \S \ref{sec:lessons} we place the future characterization of Earth-like exoplanets in the context of recent work on the frequencies and atmospheres of transiting exoplanets. We describe example experiments to test the concept of the habitable zone in \S \ref{sec:hz}. We conclude with some suggestions for future work to expand on these ideas in \S \ref{sec:conclusion}.

\section{The Systems Science Approach Reviewed} \label{sec:systems}
Most work to date has been focused on the development of a planetary systems science framework (``systems science'' hereafter for brevity) for designing and interpreting observations of potentially habitable exoplanets \citep{seager2005vegetation,meadows08,seager2010exoplanet,kaltenegger2012rocky,rugheimer2015effect,meadows16,robinson17}. The systems science framework aims to reveal the nature of individual planets based on a combination of empirical data and theoretical modeling. The empirical data should be as complete as possible to minimize model dependencies, while the theoretical models are benchmarked on the Solar System planets to maximize their accuracy \citep[e.g.,][]{robinson11,robinson14}. The ultimate goal of the systems science approach is to identify particular exoplanets that are habitable and to make statements about the possibility that they harbor life.

One strength of the systems science approach to the question of life on other planets is that a sequential roadmap can be written down that guides the construction of new facilities and the interpretation of the data they will obtain. However, the focus on individual planets in the systems science approach necessitates extensive characterization using multiple techniques and instruments, which increases the cost of the program and results in a long lead time for getting robust answers.

The systems science approach can also be interpreted to suggest that we should first characterize a small number of promising planets since that would require a smaller telescope, or that efforts to obtain more data for a small sample of planets should take precedence over a broad survey when allocating time on a larger telescope. Both of these approaches runs the risk of delivering detailed characterization of a small number of planets, but not finding any definitive indications of biosignatures and with the end result being no clearer understanding of the prevalence of life. Furthermore, the dependency on complex theoretical models will never be fully relaxed when the aim is to understand individual planets due to the limitations of astronomical remote sensing. The expected diversity of exoplanets also suggests that the models trained on the Solar System planets will be stretched in ways that may undermine their accuracy. It is telling that a range of false positive and false negative scenarios for habitability and life have already been identified \citep[e.g.,][]{wordsworth14,domogal-goldman14,reinhard17}.

We propose here to reframe the question from ``Are there other habitable or inhabited planets?'' to ``What are the frequencies of habitable and inhabited planets?''. This new question requires a larger, statistical sample and an altered observational strategy compared to what would be motivated from a simple interpretation of the systems science approach. However, we think that this question can be more robustly answered, and ultimately this information may be required to interpret the characterization of individual planets anyway.

\section{Lessons From Recent Transiting Exoplanet Studies} \label{sec:lessons}
The ongoing characterization of close-in exoplanets using transit techniques offers compelling lessons on the power of statistical comparative planetology. The \textit{Kepler} mission in particular has been transformative for not just what we know about exoplanets, but also how we go about obtaining the information. A key breakthrough in the analysis of \textit{Kepler} data was the calculation of the false positive rate for transiting planet candidates \citep{torres11,morton11}. This opened a shortcut past confirmation of individual targets, which requires a host of other observations (e.g., high resolution imaging, host star characterization, and high precision radial velocity measurements), and yielded assessments of planet frequency directly from the \textit{Kepler} data alone  \citep[e.g.,][]{fressin13,morton14,dressing15,burke15}. Besides initially requiring no additional data, another strength of the statistical approach is that it could also be extended as results from other observations became available \citep[e.g., precise host star characterization,][]{fulton17}.

The study of exoplanet atmospheres has also benefited from applying a statistical comparative planetology approach. For example, \citet{sing16} were able to show statistically that high altitude aerosols were the cause of the muted spectral features in the transmission spectra of hot Jupiters rather than low water abundances by performing a comparative study of ten planets. The water abundances determined from the individual spectra in the \citet{sing16} study had very large uncertainties, but by using the diversity of the sample the authors were able to show that low water abundances couldn't explain the observed continuum of spectral features. Another strength of the \citet{sing16} result is that it only depended on simple and generic physical models for how water abundances and aerosols affect transmission spectra.

Beyond broad statistical conclusions, comparative studies also enable the identification of outliers, which are useful for homing in on model inadequacies. For example, a study of heat transport for highly irradiated planets found that WASP-43b is a unique exception to the expectations of theoretical models and an empirical trend of heat transport vs. irradiation temperature \citep{schwartz15}. This finding has drawn attention to the possibility of clouds on the nightsides of tidally-locked planets \citep{kataria15}. Another example is the unusually high dayside albedo of Kepler-7b, which was discovered in a survey of optical secondary eclipse measurements \citep{demory11}. The properties of this planet have also motivated the development of a more comprehensive model for clouds in hot Jupiter atmospheres \citep{heng13,parmentier16}.

The statistical approach also has the benefit of enabling the combination of results from different types of observations even if they aren't targeted on the same objects. For example, \citet{schwartz15} were able to compare the geometric albedos of planets that had been studied in the optical with \textit{Kepler} to the Bond albedos of more nearby planets that had been studied in the thermal infrared with \textit{Spitzer} and \textit{Hubble}. This kind of approach has also been useful for constraining planet frequency over a broad range of parameter space by combining results from surveys that were performed with different techniques \citep[e.g.,][]{clanton17}.

\section{Worked Examples: Empirical Tests of the Habitable Zone Concept} \label{sec:hz}
One of the guiding principles in the search for other Earth-like planets is the concept of the liquid water habitable zone \citep{Kasting93,kasting14}. However, the link between the canonical habitable zone (in terms of orbital distance) and the existence of surface liquid water has not yet been shown observationally, and therefore it does not currently provide a rigorous framework for interpreting the characterization of individual planets. Testing the habitable zone hypothesis with a statistical comparative planetology approach would thus be an important step towards determining the frequency of habitable planets. 

We outline here two applications of the statistical comparative planetology approach that could be used to test for the inner and outer boundaries of the habitable zone and the prevalence of planets with temperate climates regulated by a carbonate-silicate cycle within it. We focus on how measurements of atmospheric water (H$_{2}$O) and carbon dioxide (CO$_{2}$) abundances can be used because these are likely to be the first chemical species that can be detected for terrestrial exoplanets. Water and carbon dioxide have numerous strong absorption bands throughout the optical and infrared, and thus will likely be accessible to both the transit and direct imaging approaches to studying exoplanet atmospheres. For now we remain agnostic about which technique is used. In \S\ref{sec:conclusion} we discuss how the technique being used could matter.

\subsection{Water abundances}
One possible test of the habitable zone concept is to survey the atmospheric H$_{2}$O abundances of exoplanets with a range of orbital separations. The hypothesis is that the presence of H$_{2}$O as a function of irradiation will be correlated with the boundaries of the habitable zone as predicted by models. Inside the inner edge of the habitable zone H$_{2}$O should not be abundant in the atmospheres of mature planets because it is expected to be lost to space following a runaway greenhouse process like is thought to have occurred on Venus. Similarly, beyond the outer edge of the habitable zone H$_{2}$O would be less abundant because it should freeze out.

The strengths of the statistical approach in the case of measuring atmospheric H$_{2}$O abundances are that it can likely be performed with relatively low precision per planet and that the random aspects of the planet formation process can be marginalized over. Precisely and accurately measuring the atmospheric abundance of any chemical species for an exoplanet puts stringent requirements on spectral resolution and signal-to-noise, as well as knowledge of the planet's physical and orbital properties \citep{konopacky13,kreidberg14b,lupu16,nayak17,batalha17}. Therefore, detecting the presence of H$_{2}$O on many planets may be easier to accomplish than robustly determining the atmospheric H$_{2}$O abundances for individual planets. Furthermore, determining whether an individual exoplanet has surface liquid water requires knowing not just the atmospheric H$_{2}$O abundance but also the full chemical inventory of the atmosphere so that accurate calculations of the surface temperature can be performed. Observations of a small number of planets also run the risk of encountering planets that are dry due to the stochastic nature of planet formation \citep[e.g.,][]{raymond07}.

\subsection{Carbon dioxide abundances}
The habitable zone concept rests on the assumption
of a functioning silicate-weathering feedback
\citep{Walker-Hays-Kasting-1981:negative}. This feedback should
increase the atmospheric CO$_2$ to maintain surface temperatures that
allow liquid water as a planet receives less stellar radiation. There
is some evidence that the silicate-weathering feedback has operated in
Earth history, but it is not definitive \citep{zeebe08}. A
statistical approach to exoplanet astronomy would allow us to test the
silicate-weathering feedback directly, and therefore the habitable
zone concept, by making a number of low-precision CO$_2$ measurements
on Earth-size and -mass planets inside the traditional habitable zone.
Specifically, we can use a radiative-transfer model to calculate the amount of CO$_2$
needed to maintain habitable conditions for a given stellar irradiation while marginalizing over surface temperatures and
pressures consistent with liquid water as well as uncertainties such
as clouds and other greenhouse gases. By comparing these estimates of
the amount of CO$_2$ necessary to maintain surface liquid water with
low-precision estimates of the CO$_2$ made on many planets, we can perform a statistical
test of the habitable zone concept.

\begin{figure}[t]
\begin{center}
\epsfig{file=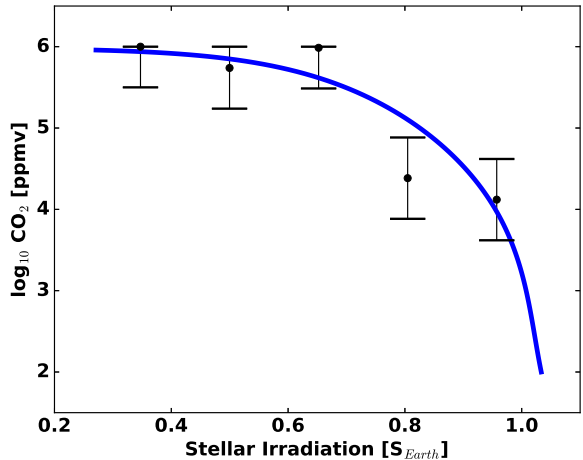, width=0.5\textwidth} 
\end{center}
\caption{This plot shows how the habitable zone concept, which assumes
  a decrease in atmospheric CO$_2$ as stellar irradiation increases, could be tested. The blue curve shows the predicted
  CO$_2$ needed to maintain a surface temperature of 290~K. The black points are binned data for hypothetical planets  that assume the
  theoretical irradiation-CO$_2$ curve but are scrambled away from it based on the
  plotted 1-$\sigma$ error bars, assuming four planets per bin. Points and error bars have a physical limit on CO$_2$ values of 10$^6$ ppmv or less. Using the error
  estimations shown here the trend predicted by the theory
  could be inferred from the data.}
\label{fig:olr_co2}
\end{figure}

Here we make a simple calculation to illustrate this methodology. A full evaluation of this idea is beyond the scope of the current work.
In Figure~\ref{fig:olr_co2} we plot in blue the theoretical prediction
for the CO$_2$ that would lead to radiative balance based on a fit to planetary albedo and infrared emission to space from a one-dimensional radiative-convective
model \citep{williams1997habitable}. We assume a saturated atmosphere with 1 bar of N$_2$, Earth-like clouds, a cosine of the solar zenith angle of 0.5, and a surface temperature of 290~K. We then create five bins of artificial data with CO$_2$  perturbed off this curve assuming Gaussian noise with a 1-$\sigma$ log error of 0.5 decades. This assumes a measurement log error of 0.5 decades; a model log error of 0.5 decades, which includes uncertainty in clouds, pressure broadening and scattering by other background gasses, and the presence of other greenhouse gases such
as CH$_4$; and four planets per bin.  Using these assumptions it would be possible to detect the downward trend in CO$_2$ as the stellar irradiation increases by measuring 20 planets. Such an inference is only possible if enough planets are measured to marginalize over the many factors other than stellar irradiation that could determine CO$_2$ even with a functioning silicate-weathering feedback. This calculation also shows that measurements on planets near Earth's irradiation (toward the inner edge of the habitable zone) will be critical for evaluating the theory because the model predicts much lower CO$_2$ values for them.

\section{Suggested Future Research} \label{sec:conclusion}
If we can step away from the idea that the only way to address the topic of habitable planets is through detailed characterization of individual objects then a comparative planetology approach could provide a useful basis for designing future experiments. We have presented some first ideas for statistical surveys that could be carried out based on a re-framing of the problem as determining the frequency of habitable worlds. Other ideas include measuring reflected stellar radiation or emitted planetary radiation to determine the planetary albedo. A statistical transition from low to high planetary albedo would represent a detection of the outer edge of the habitable zone. Another idea would be to measure the surface temperature using gap regions of the infrared spectrum. The null hypothesis would be that surface temperature would scale as irradiation to the ${\frac{1}{4}}$ power when measured for many planets. In contrast, if planetary surface temperature is regulated in the habitable zone, then it would show little dependence on stellar irradiation.

More work is needed to study the details of these ideas and those presented above. For example, beyond the outer edge of the habitable zone planets may have water in their atmospheres due to sublimation. If the water abundances are being measured through transmission spectroscopy then it may be difficult to distinguish a low stratospheric water abundance due to cold trapping from an atmosphere with a low bulk water abundance. It would also be valuable to explore whether statements could be made about the frequency of planetary inhabitance through a survey of biosignature gases such as O$_{2}$ or O$_{3}$.

A key assumption of our proposed statistical experiments is that the observations can be concentrated on terrestrial planets. This is possible in principle for transiting exoplanets, which will necessarily be orbiting M dwarfs due to their uniquely close-in habitable zones, but may be more difficult for true Earth analogues orbiting Sun-like stars, the atmospheres of which can only be probed with direct imaging and for which we will likely always lack density measurements.

The required assumption for our proposed statistical tests is an example of the more general problem that too many unknowns may make it impossible to discern the underlying trends even in a large statistical sample. Therefore it is likely that multiple types of data will need to be combined even for the statistical approach. It may also be necessary to characterize a few planets in greater detail to identify the key diagnostics for statistical investigations. Future work to flesh out the details of a statistical approach will focus on quantifying these aspects of the observing strategy. Ultimately, there is also a chance that some questions in the topic of planetary habitability and life are too complex for the statistical approach. However, we still suspect that the statistical approach will be crucial for progress in this area because getting data for more planets can help get around the unknowns, whereas it may be impossible to determine the habitability of individual exoplanets from astronomical remote sensing at high confidence.

\acknowledgments
We thank Edwin Kite, Jean-Michel D\'esert, and the anonymous referee for helpful comments on an early draft of this paper. J.L.B. thanks the members of the \textit{LUVOIR} Science and Technology Definition Team for stimulating discussions that motivated this work, and acknowledges support from the David and Lucile Packard Foundation and NASA through STScI grants GO-13021, 13467, 14792, and 14793. D.S.A. acknowledges partial support from NASA grant number NNX16AR85G, which is part of the ``Habitable Worlds'' program and from the
NASA Astrobiology Institute's Virtual Planetary Laboratory, which is supported by NASA under cooperative agreement NNH05ZDA001C. E.M.-R.K. acknowledges support from the Research Corporation for Science Advancement through the Cottrell Scholar program and from Grinnell College's Harris Faculty Fellowship.

\bibliography{ms.bib}

\end{document}